\documentclass{aa}
\usepackage{graphicx}
\usepackage[varg]{txfonts}
\usepackage{amssymb}            
\usepackage{amsfonts}           
 \usepackage{color}             
\newcommand{\be}{\begin{equation}}
\newcommand{\ee}{\end{equation}}
\newcommand{\bea}{\begin{eqnarray}}
\newcommand{\eea}{\end{eqnarray}}
\newcommand{\beq}{\begin{eqnarray}}
\newcommand{\eeq}{\end{eqnarray}}

\newcommand{\DG}{\ensuremath{^\circ}}

\graphicspath{ {./images/},{./ArtImages/},{./}}

\begin{document}

\title{Detection of spike--like structures near the front of type-II bursts}
   \author{     S. Armatas \inst{1}
                        C. Bouratzis \inst{1}
                        A. Hillaris \inst{1}
                        C.E. Alissandrakis \inst{2}
                        P. Preka-Papadema \inst{1} 
                        X. Moussas \inst{1}
                        E. Mitsakou \inst{1}
                        P. Tsitsipis \inst{3} \and      
                        A. Kontogeorgos \inst{3}}
\offprints{C. Bouratzis}
\institute{                     Department of Physics, University of Athens, 15783 Athens, Greece\\ \email{kbouratz@phys.uoa.gr}
\and                            Department of Physics, University of Ioannina, 45110 Ioannina, Greece
\and                            Department of Electronics, Technological Educational Institute of Sterea Hellas, 35100 Lamia, Greece}
\authorrunning{Armatas et al.}
\titlerunning{Spike--like structures near the front of type-II bursts}
 \date{Received .....; accepted ......}
\abstract
{}
{We examine high time resolution dynamic spectra for fine structures in type II solar radio bursts}
{We used data obtained with the {(SAO) receiver} of the Artemis-JLS (ARTEMIS-IV) solar radio spectrograph in the 450--270 MHz range at 10 ms cadence and identified more than 600 short, narrowband features. Their  characteristics, such as instantaneous relative bandwidth and total duration were measured and compared with those of spikes embedded in type IV emissions.}
{Type II associated spikes occur mostly in chains inside or close to the slowly drifting type II emission. These spikes coexist with herringbone and pulsating structures. Their average duration is 96\,ms and their average relative bandwidth 1.7\%. These properties are not different from those of type IV embedded spikes. It is therefore possible that they are signatures of small-scale reconnection along the type II shock front.}
{}
   \maketitle
%
\section{Introduction}\label{Intro}
Solar type II radio bursts are thought to be signatures of either {shocks driven by coronal mass
ejections (CME)} or flare blast shock waves \citep{Vrsnak08,Pick08}.
The radio emission is the result of energetic electrons accelerated by the shock. On the dynamic spectra they appear as slowly drifting bands, dubbed backbone, from high to low frequencies \citep{Roberts59,Krueger1979}.  {The type II radio bursts 
often exhibit} a characteristic fundamental-harmonic (F-H) structure \citep{Maxwell1962} and sometimes show a split in two lanes  by a small frequency offset of \mbox{$\approx$f/8-f/4} \citep{Smerd75, Vrsnak04}.

Type II bursts exhibit fine structure. A common feature is the so-called herringbone structure, which is attributed to shock-accelerated electrons producing narrow bandwidth, type III-like groups, originating at the backbone and drifting toward higher and lower frequencies \citep{Roberts59,Cairns1987}. Slow drift fiber-like structures were reported by \citet{Chernov1997} and \citet{Chernov2007b} and interpreted as whistler wave packets upstream of the shock front. Narrowband drifting fibers, thought to be the result of type II emission refracted on inhomogeneous structures of the CME preceding the shock, have also been reported by \citet{2009_Afanasiev}. 

Spike bursts, which are abundant in type IV emissions ({e.g.,} Bouratzis et. al. 2016), are conspicuously absent from the type II fine structure, apart from one report \citep{chernov2016} of spike-like structures in a single decametric event. The spike bursts have very short duration ($\sim$ 100 ms in the metric frequency range) and bandwidth ($\sim$ 1-2\%) and are thought to be the signature of small-scale electron acceleration events through magnetic reconnection \citep{Nindos07}.

We report the detection of narrowband spike-like structures, which are shorter than 100 ms and embedded in type II burst harmonic emission, in observations obtained with the high time resolution (10 ms) Artemis-JLS {acousto-optic spectrograph (SAO) receiver}  in the 450-270 MHz range. In section \ref{InstrObs} we describe our observations and data reduction followed by an overview of the events selected for study (section \ref{OverView}) and the presentation of their fine structure (section \ref{FS}). Results are summarized and discussed in section \ref{DisC}.

\begin{figure*}
\begin{center}
\includegraphics[width=0.95\textwidth]{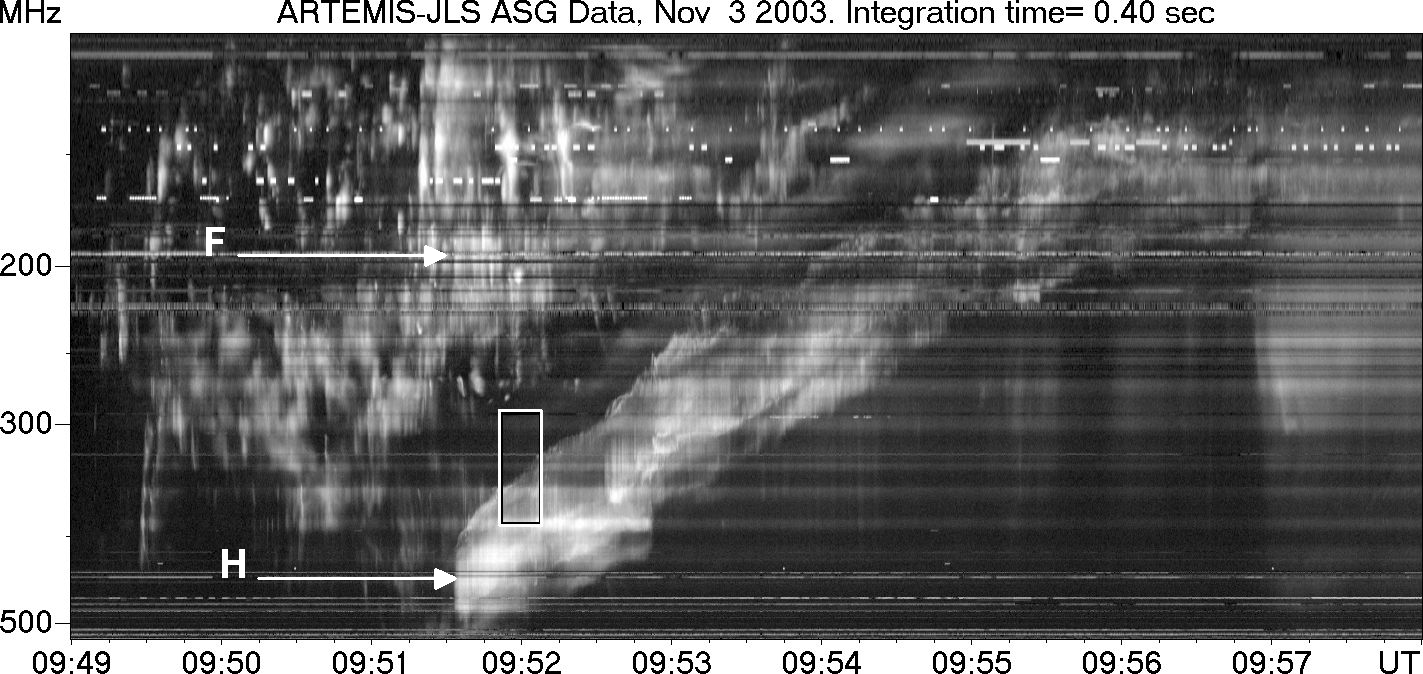}
\end{center}
\caption{Dynamic spectrum of the complex first event (\mbox{SOL2003-11-03T09:43:20}), observed with the Artemis-JLS/ASG receiver. The box indicates the segment discussed in Sect. \ref{FS} and the arrows the fundamental (F) and harmonic (H) lanes.}
\label{03B03ASG5Channel}
\end{figure*}

\begin{figure*}
\begin{center}
\includegraphics[width=0.95\textwidth]{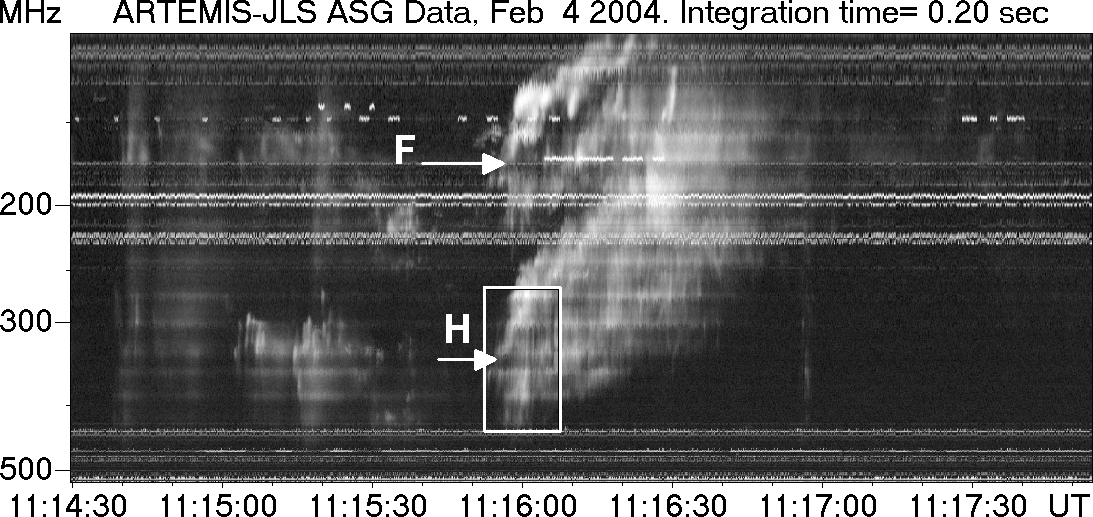}
\end{center}
\caption{Same as Fig. \ref{03B03ASG5Channel} for the second event, SOL2004-02-04T11:12:00.} 
\label{04204}
\end{figure*}
\begin{figure*}
\begin{center}
\includegraphics[width=0.95\textwidth]{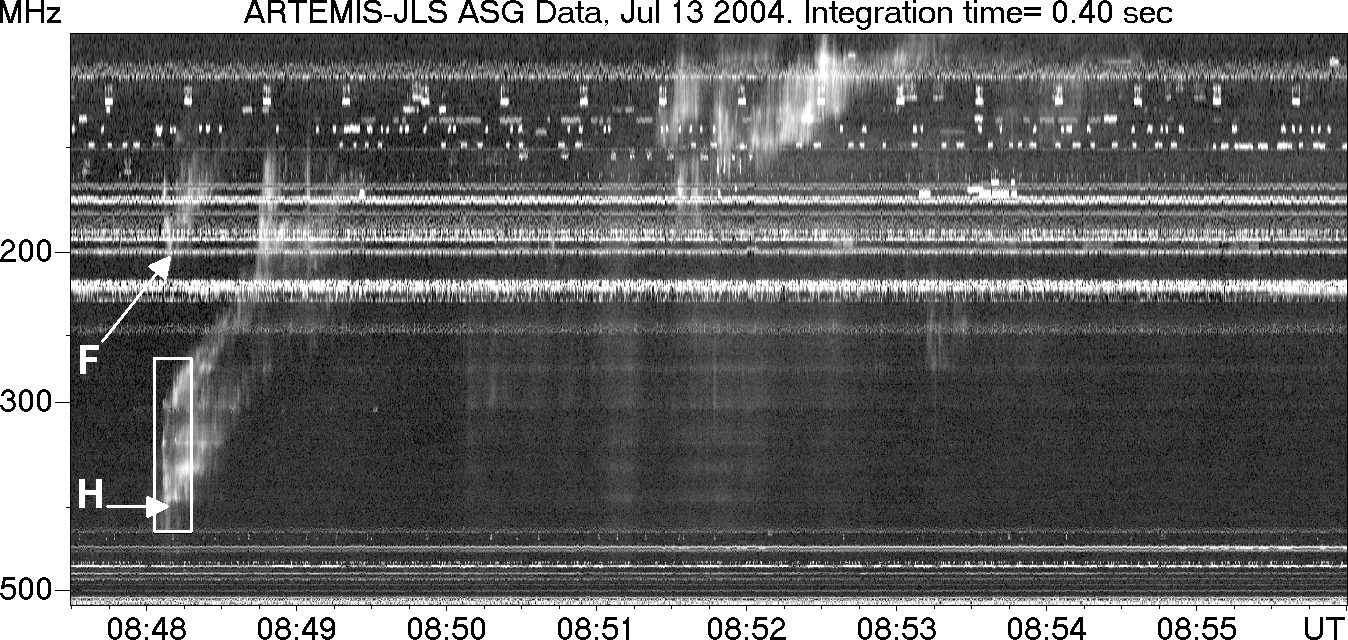}
\end{center}
\caption{Same as Fig. \ref{03B03ASG5Channel}, for the third event, SOL2004-07-13T08:40:00.}
\label{20040713s}
\end{figure*}
\begin{figure*}
\begin{center}
\includegraphics[width=0.95\textwidth]{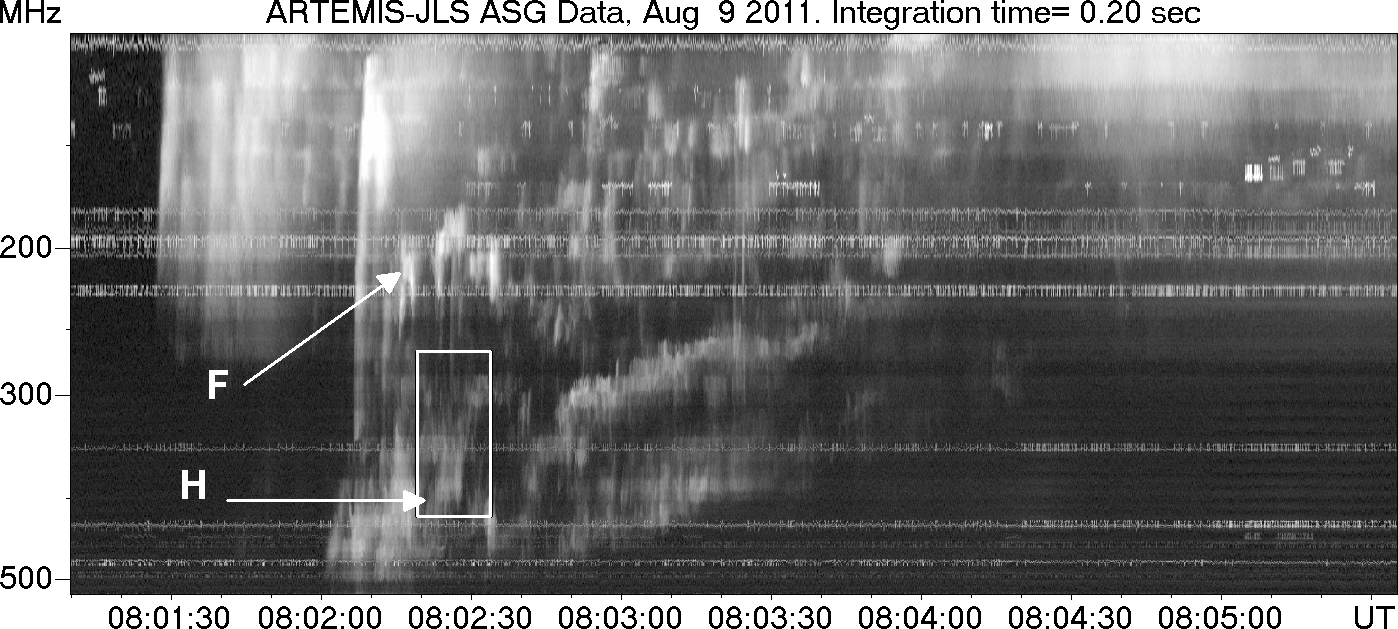}
\end{center}
\caption{Same as Fig. \ref{03B03ASG5Channel} for the fourth event, SOL2011-08-09T07:48:00.}
\label{20110809}
\end{figure*}
\section{Observations and data reduction} \label{InstrObs}

The Artemis-JLS (ARTEMIS-IV) solar radio spectrograph has been operating at Thermopylae since 1996 \citep{Caroubalos01,Kontogeorgos06}. The observations cover the frequency range from 20 to 650 MHz. The spectrograph has a 7 m moving parabola fed by a log-periodic antenna for the 100-650 MHz range and a stationary inverted V fat dipole antenna for the 20-100 MHz range. Two receivers operate in parallel: a sweep frequency analyzer (ASG) covering the 650-20 MHz range in 630 channels with a cadence of  ten samples/sec and a low-noise, high sensitivity multi-channel acousto-optical analyzer (SAO), which covers the 270-450 MHz range in 128 channels with a time resolution of 10 ms. 

Information about the corresponding flares such as location, class, and importance were obtained from the { National Oceanic and Atmospheric Administration (NOAA) catalog}. Information about associated CMEs { was acquired from the Large Angle and Spectrometric Coronagraph (LASCO) on board the Solar and Heliospheric Observatory (SoHO) mission} \citep{Brueckner95} and the CME catalog \citep{Gopalswamy09}.
{From a large number of type II bursts observed with Artemis-JLS/ASG in the 1998-2013 period we selected four events that were well covered by SAO receiver at frequencies above 270 MHz. All flares associated with these events occurred at longitudes \mbox{$\ge$50$\DG$}.}

\section{Overview of the selected events}\label{OverView}%

The first event (2003 November 3, SOL2003-11-03T09:43:20) was related to a GOES X3.9 class flare at 08N\DG\,77W\DG\ in AR 10488 near the west limb. The flare started at 09:43 UT and reached peak intensity at 09:55 UT. A CME was first detected at 10:06 UT with a backward--extrapolated lift-off at 09:53 UT. 

Fig.~\ref{03B03ASG5Channel} shows  an expanded view of event observed with the ASG receiver of Artemis-JLS. The dynamic spectrum shows the type II burst, {which has a relative frequency drift  \mbox{df/fdt=0.005\,$\rm{s}^{-1}$}}, and associated type III and IV radio bursts. During the rise phase of the flare, a group of type III bursts overlap the fundamental emission  of the type II burst in the 270 to 30 MHz range. The type II harmonic starts at 09:51:30 UT and 520 MHz, mostly within the Artemis-JLS/SAO frequency range.

The second event (2004 February 04, SOL2004-02-04T11:12:00) was associated with a C9.9 GOES flare, which occurred at 11:12 UT,  07$\DG$\,S49$\DG$W (AR 10547), and a CME, which was observed by LASCO at 11:54 UT (backward--extrapolated lift-off time 11:19:28 UT). Fig. \ref{04204} presents an overview of the type II  F-H burst  with a relative drift rate of \mbox{3.8\,$\rm{s}^{-1}$} and the associated  activity which includes the precursor of the type II burst \citep{Klassen03}.

The third event (2004 July 13, SOL2004-07-13T08:40:00) was related to a GOES M5.4 flare at 12$\DG$\,N52$\DG$W (AR 10646) and a halo CME. The flare started at 08:40 UT and reached its peak intensity at 08:48 UT, while the CME lift-off was at 08:46 UT. Spectral radio observations from ARTEMIS-JLS/ASG in the 20-500 MHz range are presented in Fig.\ref{20040713s}. A small group of type III bursts (08:46-08:47 UT, not shown in the figure) preceded the type II burst, which appears at 08:48 UT and exhibits F-H structure. The starting frequency of the fundamental lane is 187\,MHz and has a relative frequency drift rate of 0.011\,s$^{-1}$.

The fourth event (August 9 2011, SOL2011-08-09T 07:48:00) was associated with an X6.9 class flare from AR 11263 (N17$\DG$\,W69$\DG$), starting at 07:48 UT (08:05 peak) and a fast (1610\,km/s) CME recorded by SoHO/LASCO at 08:12 UT with extrapolated lift-off at 07:52 UT. On the dynamic spectrum (Fig. \ref{20110809}) we have a group of type III bursts before the type II (F-H) and another III/U burst group starting near the onset time and frequency of the type II. The latter has a relative  drift rate of 0.005\,$\rm{s}^{-1}$.

\begin{figure*}
\begin{center}
\includegraphics[width=.95\textwidth]{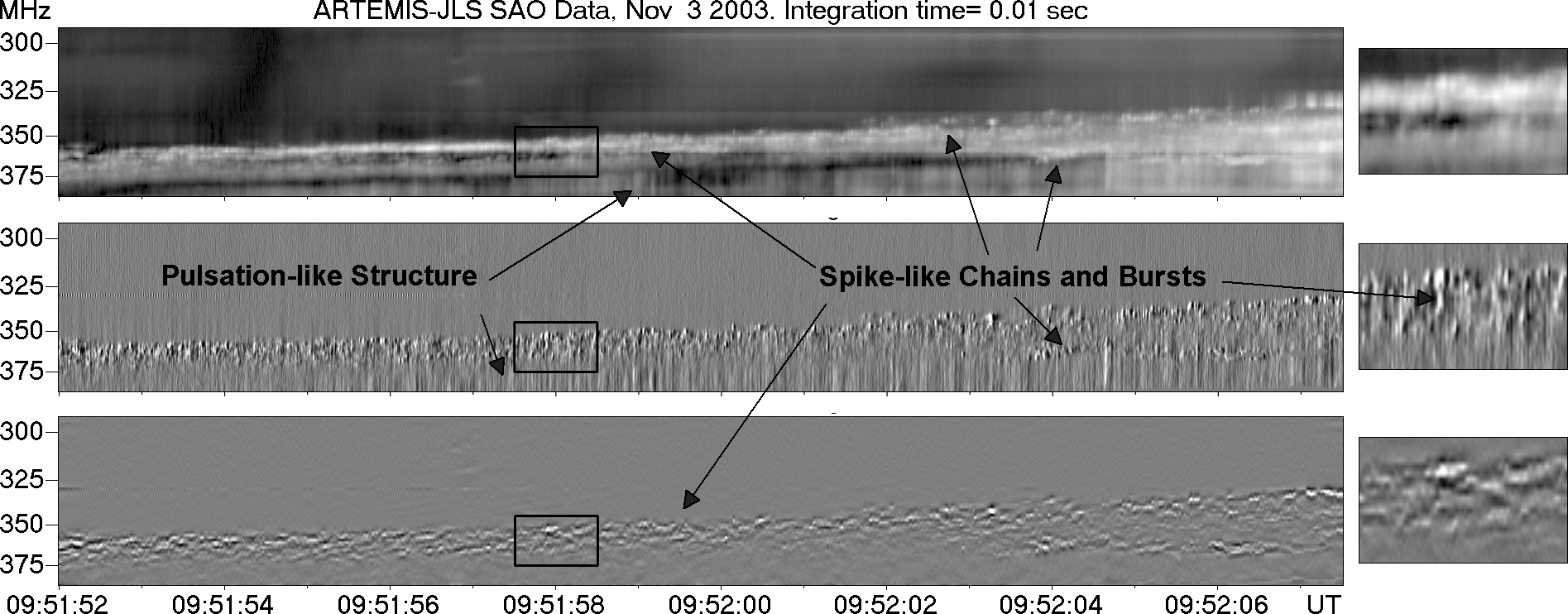}
\end{center}
\caption{Part of the high time resolution dynamic spectrum for the first event. The top panel shows the original spectrum, the middle panel the derivative with respect to time (differential spectrum), and the lower panel the spectrum after high pass filtering in time and frequency. The boxes indicate a 1\,s by 30\,MHz segment of the spectrum that is shown enlarged on the right. 
The arrows point at type II associated fine structures.} 
\label{20031103FS}
\end{figure*}
\begin{figure*}
\begin{center}
\includegraphics[width=.95\textwidth]{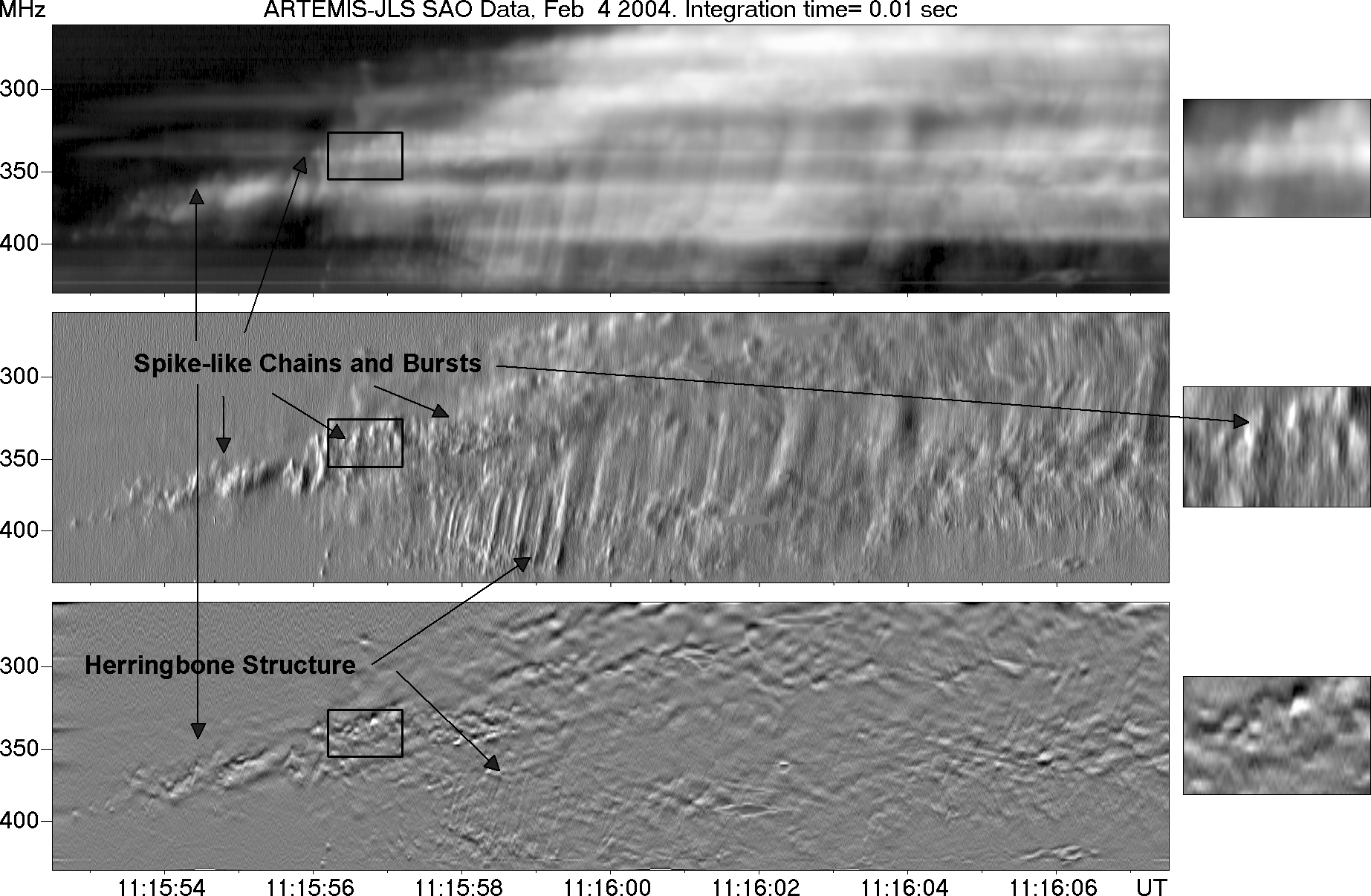}
\end{center}
\caption{Same as Fig. \ref{20031103FS} for the second event.}
\label{20040204FS}
\end{figure*}

\begin{figure*}
\begin{center}
\includegraphics[width=.95\textwidth]{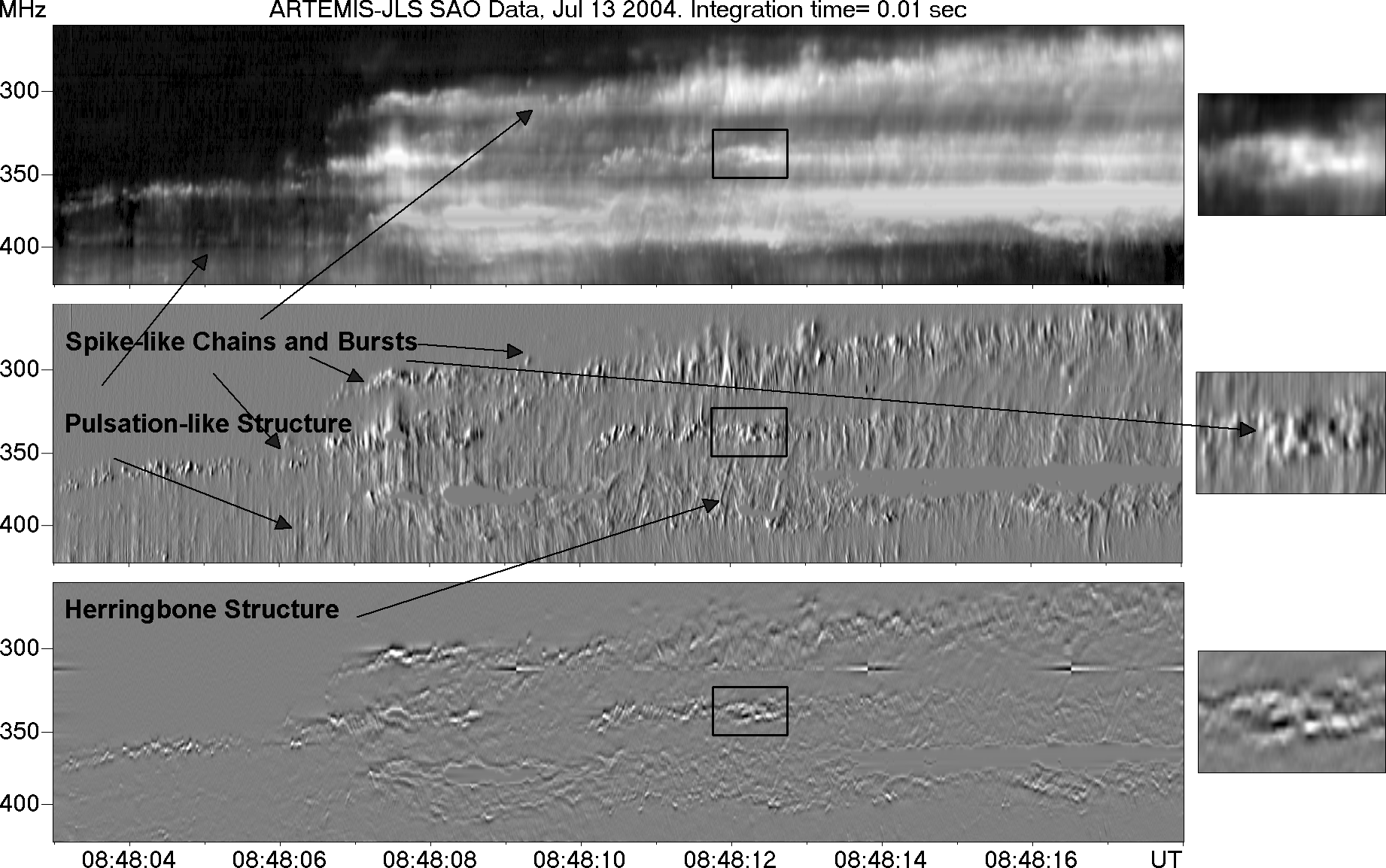}
\end{center}
\caption{Same as Fig. \ref{20031103FS} for the third event. The uniform gray patches in the middle of the differential spectrum are due to saturation effects. Artifacts near 312\,MHz in the filtered spectrum are due to interference signals.} 
\label{20040713FS}
\end{figure*}

\begin{figure*}
\begin{center}
\includegraphics[width=.95\textwidth]{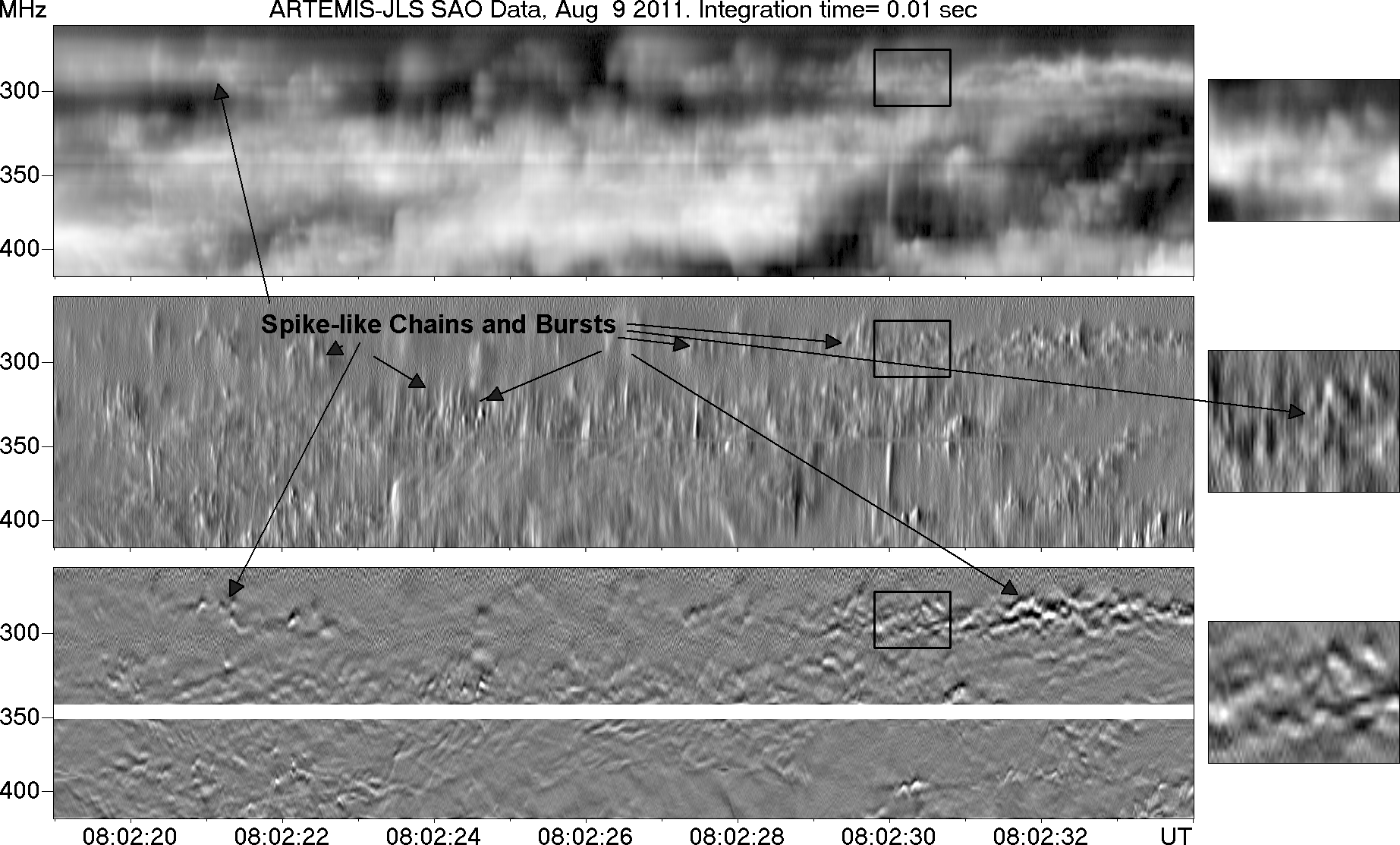}
\end{center}
\caption{Same as Fig. \ref{20031103FS} for the fourth event. Some frequency channels around 350\,MHz were strongly affected by radio interference and have been deleted in the filtered spectrum.}
\label{20110809FS}
\end{figure*}

\section{Fine structure} \label{FS}
Fig. \ref{20031103FS} shows a 15 s time interval of the first event (box in Fig.\ref{03B03ASG5Channel}) with10\,ms time resolution, which includes the shock front of the type II harmonic emission. In this and other events, { the fundamental emission lane was outside of the SAO frequency range}. In addition to the original spectrum, we give its derivative with respect to time (differential spectrum) that enhances the fine structure and is less sensitive to receiver gain variations and radio interference signals. We also give the spectrum after the application of a high pass Gaussian filter with a width of 0.6\,s by 5\,MHz that suppressed large-scale structure; {cf.} \cite{Bouratzis2016}. 

We note a wealth of fine structure. The most prominent are chains of dot-like features resembling spikes commonly seen in type IV bursts. These chains are best visible in the filtered spectrum and are embedded in emission lanes, which drift almost parallel to one another and probably correspond to different regions of the expanding shock. In addition, we see high-drift structures before and during the chains, which are best visible in the differential spectrum. They are nearly vertical in the dynamic spectrum, hence they look more like pulsations rather than herringbones. These pulsation-like structures appear at frequencies higher than the type II lanes, drifting toward lower frequencies and seem to terminate at the spike chains.

In Fig. \ref{20040204FS}, we give  a  15 s segment of the second event fine structure. This structure contains herringbones, chains of spike-like structures, and narrowband bursts of the type III family, and some of these appear in the gap between the two emission lanes. We note that a spike chain starts $\sim3$\,s before the main type II. The herringbones have a relative frequency drift rate of \mbox{0.55$\rm{s}^{-1}$} (speed $\approx 0.25~\rm{c}$  under the assumption of fourfold \citet{Newkirk} coronal density-height variation). The herringbone speed is consistent with published results { by \citet{Cairns1987}, \citet{Mann05}, and \citet{2015Carley}.}

{The fine structure is similar in the third event, of which a 15 s dynamic spectrum is shown in Fig. \ref{20040713FS}. In this case the harmonic of the type II is almost entirely within the SAO frequency range and spike-like structures as well as herringbones were also recorded at the low frequency side of the type II (H).  In this case again we have a spike chain before the start of the main type II. 

A  15 s segment of the dynamic spectrum of the fourth event is given in  Fig. \ref{20110809FS}. There are narrowband, spike-like structures in drifting chains during the shock and a more diffuse group at the onset of the Type II. 

On the 10 ms SAO dynamic spectra we measured the instantaneous relative bandwidth ($\delta f/f$) and the total duration ($\delta\tau$) of 642 narrowband bursts. The identification of individual bursts was done by visual inspection, and the bandwidth and duration were measured after fitting the temporal and spectral profiles with a smooth curve as in \citet[{Fig. 3}]{Bouratzis2016}.  In Table \ref{TableSpikes} we summarize the average parameters of the  type II-associated spike-like structures. These parameters are very close to the corresponding parameters of type IV-associated spikes from \citet[][their Table 1]{Bouratzis2016}, which are also included in the table for comparison. 

\begin{table}
\caption{ Parameters of type II-associated spike-like bursts and type IV-associated spikes from \citet{Bouratzis2016}. }
\label{TableSpikes}
\begin{tabular}{lrr}
\hline
Parameter                               & Type II spike-like& Type IV spikes    \\
                                                & bursts                        &                                       \\                                              
\hline                                                                                              
Number of events                & 642                           & 11579                          \\
\hline
Duration (ms)                   &                                       &                                         \\ 
Average                                 & 96                            & 100                             \\
{ Standard deviation}   & 54                            & 66                            \\  
\hline                                  
Relative bandwidth              &                                       &                                         \\ 
Average                                 & 1.7\%                         & 2.0\%                           \\                                              
{    Standard deviation}& 0.5\%                         & 1.1\%                         \\                                                                                                                                                 
\hline
\end{tabular}
\end{table}

The duration--frequency and bandwidth--frequency dependence of the spikes (narrowband structures) are usually expressed by phenomenological power laws of the form $\delta \tau \propto  f^{-a}$, ($a \approx 1.32$) \citep{Guedel1990, Rozhansky2008,Sirenko2009}, and $\delta f \propto 0.66f^{0.42}$ by \citep{Csillaghy1993}, respectively (see plots in Fig. 6 of Bouratzis et. al., 2016}). The present measurements are consistent with these empirical relations.
 
\section{Summary and discussion} \label{DisC}

We observed a large number of spikes in metric type II radio bursts, using the SAO receiver of the Artemis-JLS radio spectrograph in the frequency range of 450-270 MHz. Their detection and the measurement of their duration and bandwidth was made possible thanks to the high sensitivity and time resolution of the instrument. As mentioned in the Introduction, \citet{chernov2016} reported  fine structure, including spike-like structures, in a single decametric type II event. The author did not, however,  elaborate further or provide any parameters of the observed structures to judge if these are similar to the spikes reported in this work. 
 
Our comparison of the duration and bandwidth of type II associated spikes with those of type IV spikes (Table \ref{TableSpikes}) show that they are very similar. The duration is comparable, while the difference in bandwidth is probably justified by the fact that its measurement depends both on the radio source size and the ambient density gradient, which is expected to vary considerably between disturbed plasma of the shock front and the type IV environment.

 The vast majority of the spikes detected were aligned in chains along the type II lanes. The presence of spike-like fine structure in or near the type II lanes is probably a signature of non-thermal electrons originating in small-scale acceleration episodes. A possible candidate is an ensemble of reconnection events between the magnetic field downstream of the shock and the ambient magnetic field or preexisting magnetic structures.
  
In the future we expect to combine our spectral observations with imaging observations from the Nan\c cay Radio Heliograph (NRH) to check the position of spikes with respect to that of the type II lanes. We will also examine our type II collection observed with the 100\,ms resolution ASG receiver of Artemis-JLS for traces of spike structures in an effort to expand our sample of events.

\begin{acknowledgements} 
The authors thank C. Karaberis for his assistance with the data reduction of the Artemis-IV/JLS observations.  We also wish to thank the Onassis Foundation for financial support (Grant 15153) for the continued operation of the ARTEMIS-JLS radio spectrograph and the University of Athens Research Committee for Grant 15018. The anonymous reviewer provided useful suggestions, which have improved the quality of this article. 
\end{acknowledgements}
\bibliographystyle{aa}
\bibliography{Ref/P01,Ref/P02,Ref/P03,Ref/P04,Ref/P05,Ref/P06,Ref/P07,Ref/P08,Ref/BurstsII,Ref/BurstsIII,Ref/General,Ref/SEE2007,Ref/Books,Ref/Density,Ref/Spikes,Ref/Temp,Ref/FSCatalogue}

\begin{thebibliography}{26}
\expandafter\ifx\csname natexlab\endcsname\relax\def\natexlab#1{#1}\fi

\bibitem[{{Afanasiev}(2009)}]{2009_Afanasiev}
{Afanasiev}, A.~N. 2009, Annales Geophysicae, 27, 3933

\bibitem[{{Bouratzis} {et~al.}(2016){Bouratzis}, {Hillaris}, {Alissandrakis},
  {Preka-Papadema}, {Moussas}, {Caroubalos}, {Tsitsipis}, \&
  {Kontogeorgos}}]{Bouratzis2016}
{Bouratzis}, C., {Hillaris}, A., {Alissandrakis}, C.~E., {et~al.} 2016, \aap,
  586, A29

\bibitem[{{Brueckner} {et~al.}(1995){Brueckner}, {Howard}, {Koomen},
  {Korendyke}, {Michels}, {Moses}, \& \etal}]{Brueckner95}
{Brueckner}, G.~E., {Howard}, R.~A., {Koomen}, M.~J., {et~al.} 1995, \solphys,
  162, 357

\bibitem[{{Cairns} \& {Robinson}(1987)}]{Cairns1987}
{Cairns}, I.~H. \& {Robinson}, R.~D. 1987, \solphys, 111, 365

\bibitem[{{Carley} {et~al.}(2015){Carley}, {Reid}, {Vilmer}, \&
  {Gallagher}}]{2015Carley}
{Carley}, E.~P., {Reid}, H., {Vilmer}, N., \& {Gallagher}, P.~T. 2015, \aap,
  581, A100

\bibitem[{{Caroubalos} {et~al.}(2001){Caroubalos}, {Maroulis}, {Patavalis},
  {Bougeret}, {Dumas}, {Perche}, \& \etal}]{Caroubalos01}
{Caroubalos}, C., {Maroulis}, D., {Patavalis}, N., {et~al.} 2001, Experimental
  Astronomy, 11, 23

\bibitem[{Chernov(2016)}]{chernov2016}
Chernov, G. 2016, Solar Flares: Investigations and Selected Research (Nova
  Science Publishers, Inc.), iSBN: 978-1-53610-204-8

\bibitem[{{Chernov}(1997)}]{Chernov1997}
{Chernov}, G.~P. 1997, Astronomy Letters, 23, 827

\bibitem[{{Chernov} {et~al.}(2007){Chernov}, {Stanislavsky}, {Konovalenko},
  {Abranin}, {Dorovsky}, \& {Rucker}}]{Chernov2007b}
{Chernov}, G.~P., {Stanislavsky}, A.~A., {Konovalenko}, A.~A., {et~al.} 2007,
  Astronomy Letters, 33, 192

\bibitem[{{Csillaghy} \& {Benz}(1993)}]{Csillaghy1993}
{Csillaghy}, A. \& {Benz}, A.~O. 1993, \aap, 274, 487

\bibitem[{{Gopalswamy} {et~al.}(2009){Gopalswamy}, {Yashiro}, {Michalek},
  {Stenborg}, {Vourlidas}, {Freeland}, \& {Howard}}]{Gopalswamy09}
{Gopalswamy}, N., {Yashiro}, S., {Michalek}, G., {et~al.} 2009, Earth Moon and
  Planets, 104, 295

\bibitem[{{Guedel} \& {Benz}(1990)}]{Guedel1990}
{Guedel}, M. \& {Benz}, A.~O. 1990, \aap, 231, 202

\bibitem[{{Klassen} {et~al.}(2003){Klassen}, {Pohjolainen}, \&
  {Klein}}]{Klassen03}
{Klassen}, A., {Pohjolainen}, S., \& {Klein}, K.-L. 2003, \solphys, 218, 197

\bibitem[{{Kontogeorgos} {et~al.}(2006){Kontogeorgos}, {Tsitsipis},
  {Caroubalos}, {Moussas}, {Preka-Papadema}, {Hilaris}, \&
  \etal}]{Kontogeorgos06}
{Kontogeorgos}, A., {Tsitsipis}, P., {Caroubalos}, C., {et~al.} 2006,
  Experimental Astronomy, 21, 41

\bibitem[{{Krueger}(1979)}]{Krueger1979}
{Krueger}, A. 1979, Geophysics and Astrophysics Monographs, 16

\bibitem[{{Mann} \& {Klassen}(2005)}]{Mann05}
{Mann}, G. \& {Klassen}, A. 2005, \aap, 441, 319

\bibitem[{{Maxwell} \& {Thompson}(1962)}]{Maxwell1962}
{Maxwell}, A. \& {Thompson}, A.~R. 1962, \apj, 135, 138

\bibitem[{{Newkirk}(1961)}]{Newkirk}
{Newkirk}, G.~J. 1961, \apj, 133, 983

\bibitem[{{Nindos} \& {Aurass}(2007)}]{Nindos07}
{Nindos}, A. \& {Aurass}, H. 2007, in Lecture Notes in Physics, Vol. 725, The
  High Energy Solar Corona: Waves, Eruptions,Particles, Berlin Springer Verlag,
  ed. {K.-L.~Klein \& A.~L.~MacKinnon}, 251--277

\bibitem[{{Pick} \& {Vilmer}(2008)}]{Pick08}
{Pick}, M. \& {Vilmer}, N. 2008, \aapr, 16, 1

\bibitem[{{Roberts}(1959)}]{Roberts59}
{Roberts}, J.~A. 1959, Australian Journal of Physics, 12, 327

\bibitem[{{Rozhansky} {et~al.}(2008){Rozhansky}, {Fleishman}, \&
  {Huang}}]{Rozhansky2008}
{Rozhansky}, I.~V., {Fleishman}, G.~D., \& {Huang}, G.-L. 2008, \apj, 681, 1688

\bibitem[{{Sirenko} \& {Fleishman}(2009)}]{Sirenko2009}
{Sirenko}, E.~A. \& {Fleishman}, G.~D. 2009, Astronomy Reports, 53, 369

\bibitem[{{Smerd} {et~al.}(1975){Smerd}, {Sheridan}, \& {Stewart}}]{Smerd75}
{Smerd}, S.~F., {Sheridan}, K.~V., \& {Stewart}, R.~T. 1975, \aplett, 16, 23

\bibitem[{{Vr{\v s}nak} \& {Cliver}(2008)}]{Vrsnak08}
{Vr{\v s}nak}, B. \& {Cliver}, E.~W. 2008, \solphys, 253, 215

\bibitem[{{Vr{\v s}nak} {et~al.}(2004){Vr{\v s}nak}, {Magdaleni{\'c}}, \&
  {Zlobec}}]{Vrsnak04}
{Vr{\v s}nak}, B., {Magdaleni{\'c}}, J., \& {Zlobec}, P. 2004, \aap, 413, 753

\end{thebibliography}

\end{document}